\documentclass[aps,prb,floatfix,twocolumn,showpacs]{revtex4}

\usepackage{epsfig}
\usepackage{amsmath,amssymb}
\begin{document}

\title{Theory of digital magneto resistance in ferromagnetic resonant tunneling diodes}

\author{Christian Ertler\footnote{email:christian.ertler@physik.uni-regensburg.de}}
\author{Jaroslav Fabian\footnote{email:jaroslav.fabian@physik.uni-regensburg.de}}
\affiliation{Institute of Theoretical Physics, University of
Regensburg, Universit\"atsstrasse 31, D-93040 Regensburg, Germany}

\begin{abstract}
We propose a ferromagnetic spintronic system, which consists of two serial connected resonant
tunneling diodes. One diode is
nonmagnetic whereas the other comprises a ferromagnetic emitter and
quantum well. Using a selfconsistent coherent transport
model we show that the current-voltage characteristic of the ferromagnetic diode
can be strongly modulated by changing the relative orientation of
the magnetizations in the emitter and quantum well, respectively. By a continuous change of
the relative magnetization angle the total resistance exhibits a
discrete jump realizing digital magneto
resistance. The interplay between  the emitter's Fermi energy level and the relative magnetization
orientations allows to tailor the current voltage characteristics of the ferromagnetic
diode from ohmic to 
negative differential resistance regime at low voltages. 

\end{abstract}

\maketitle

\section{Introduction}

The development of ferromagnetic dilute magnetic semiconductors
(DMSs) \cite{Oh98,Di02,Dietl:book06, PeAbNo03,LiYuMo05,JuSiMa06} has opened the possibility of
novel all semiconductor spintronic device concepts, in which the charge current can be
modulated by the carriers spin.\cite{ZuFaSa04} For example, spin
dependent resonant tunneling  have been investigated both
experimentally and theoretically in magnetic double barrier
heterostructures with either a ferromagnetic quantum well (QW), e.g., GaMnAs,
\cite{PeDeCh00,HaTaAs00,OiMoKa04,FuWoLi04,LeKu05,GaReBa05,OhHaTa05,MaElGe05}
or a paramagnetic QW, which exhibits a giant $g$-factor, e.g.,
ZnMnSe.\cite{GrKeFi01,SlGoSl03,BeBeBo05} By employing resonant
interband tunneling an effective injection of spin-polarized
electrons into nonmagnetic semiconductors has been demonstrated.
\cite{PeDeCh03,VuMe03} Moreover, enhanced tunneling
magnetoresistance (TMR) has been predicted and found in double
barrier magnetic tunneling junctions, \cite{PeChDe02,HaTaAs00,OhHaTa05,MaElGe05} 
in which a nonmagnetic QW is sandwiched between two magnetic electrodes.
Recently, high magnetocurrents [relative current magnitudes for parallel (P) and
antiparallel (AP) orientations of the magnetizations] have been predicted in two coupled
magnetic QWs.\cite{ErFa06b}

Conventional nonmagnetic resonant tunneling diodes (RTDs) are technologically interesting due to
their extreme high speed and low power performance. They allow
for novel circuit concepts based on their specific negative differential resistance 
(NDR) behavior.\cite{MaFo:book03} A logic gate named MOBILE (MOnostable-BIstable
Transition Logic Element), which consists of
two serial connected RTDs, a load and driver, has been 
proposed and realized by Maezawa and Mizutani.\cite{MaMi93,MaAkMi94} The device
is driven by an oscillating input voltage, which performs the
transition between the mono- and bistable working point regimes. At
low input voltages Kirchoff's laws allow for only one stable dc
working point. However, for high input voltages two stable working
points become possible due to the N-shaped current voltage (IV) charateristic
of both the load and
the driver RTD. Which of the two working points is actually realized
depends on the difference of the load and driver peak currents. When
the load peak current is higher than the driver one the working
point voltage is high and vice versa. The whole device works
actually as a comparator of the load and driver's peak current. For
a detailed discussion of the operation principle see Refs.
\onlinecite{MaMi93,MaAkMi94,MaFo:book03}.

We have recently proposed \cite{ErFa06a} that by replacing the driver
by a magnetic RTD with a paramagnetic QW the circuit exhibits 
what we call digital magneto resistance
(DMR): the output voltage jumps from low to high after the mono-to-bistable
transition if the external magnetic field, which controls the Zeeman
splitting in the QW, is higher than some threshold value. The
threshold value of the magnetic field can be controlled by a gate
voltage, which influences the peak current of the load. The proposed
device performs a direct digital conversion of an analog magnetic
signal and might be used as a fast magnetic read head. DMR has actually been
experimentally demonstrated in an earlier setup by shunting a metallic giant
magnetoresistance (GMR) element to a nonmagnetic driver RTD,\cite{HaMaCh01} 
having the advantage of being nonvolatile upon the loss 
of power: the state of the device is stored in the magnetization
direction of a particular ferromagnetic layer of the GMR-element.
Such nonvolatile devices are attractive for fast and reliable data
storage, e.g, in random access memory applications \cite{Da99} or for
reprogrammable logics,\cite{Wh98} in which the logical function of
the circuit can be tuned by changing the magnetic state of the
device.

In this paper we lay down the physical principles of a 
nonvolatile ferromagnetic MOBILE. We propose to use
a driver RTD, which comprises a ferromagnetic emitter and QW. By
performing realistic selfconsistent calculations of the
IV-characteristics DMR is observed when the QW magnetization is
tilted. For a proper choice of the emitter's Fermi energy the
driver IV can be changed from ohmic to negative differential
resistance behavior in the low voltage regime depending on the relative magnetization
orientation. 

The paper is organized as follows. A discussion of the system 
and the selfconsistent transport model is presented in section~\ref{physmod}. 
The simulation results are shown
and discussed in detail in section~\ref{simres}, and, finally, conclusions are
given in section~\ref{conclus}.


\section{\label{physmod}  Model}

The operation of the magnetic MOBILE is based on the
change of the driver's peak current by applying an external magnetic
field or by changing some magnetization direction in the device. In
order to realize a nonvolatile ferromagnetic MOBILE one can use
 the double barrier TMR structure,\cite{PeChDe02} in which a
nonmagnetic QW is sandwiched between two ferromagnetic leads. For
low voltages the structure yields large magnetocurrents, but for higher
voltages, when the exchange splitting of the collector lead is
shifted far below the band edges of the emitter lead by the applied
voltage, the magnetocurrent becomes small. This means that for
an exchange splitting of the order of a few tens of meV the peak
current is hardly influenced by the relative orientation of
collector's magnetization, since the peak voltage is already too
high. Hence, hardly any DMR would be observed in such a TMR-device.

Another setup to realize non-volatility would be to use a
ferromagnetic QW instead of a paramagnetic one. In the proposed
paramagnetic MOBILE,\cite{ErFa06a} the driver peak current decreases
with increasing Zeeman splitting. In ferromagnetic
materials it is easier to change the orientation than the
magnitude of the magnetization. If the exchange splitting is
strongly anisotropic a nonvolatile MOBILE could indeed be realized
by a ferromagnetic QW and by rotating its magnetization direction.
However, in an isotropic case as considered here an additional ferromagnetic lead is
necessary to observe the DMR-effect. We consider a ferromagnetic
emitter, since a ferromagnetic collector lead would again have
little influence on the peak current.

\begin{figure}
\centerline{\psfig{file=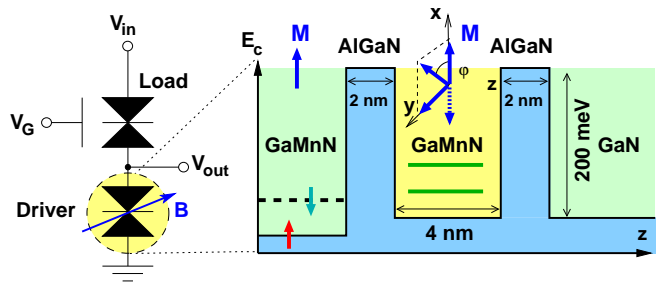,width=1\linewidth}}
\caption{(Color online) Left: The circuit configuration of the
proposed ferromagnetic MOBILE. The load is a conventional RTD, whose peak
current can be modified by an external gate voltage $V_G$. The
driver device consists of a ferromagnetic RTD. The peak current of the driver is controlled by
twisting the QW magnetization. Right: The schematic conduction band profile of the
ferromagnetic RTD [here made of a Ga(Al,Mn)N
material system] used in the numerical
simulations discussed in the text.}\label{fig:scheme}
\end{figure}

In order to investigate the characteristic physical effects of such
a structure as shown in Fig.~\ref{fig:scheme}, we consider the emitter and QW to be made of a
generic, moderately doped n-type ferromagnetic
semiconductor. We also expect to observe similar effects in a p-type
ferromagnetic semiconductor, since in the simplest approach the
heavy and light holes can be treated by an effective mass
model,\cite{MeWaRi85} analogous to our description of the conduction
electrons here. In todays p-type ferromagnetic DMSs, e.g., GaMnAs
the Fermi energy cannot be chosen freely, since the ferromagnetic
order appears only at high hole densities.\cite{DiOhMa01} However, for
device applications a decoupling of the ferromagnetic order and the
doping would be advantageous. There have been several
experimental reports on moderately doped n-type ferromagnetic
semiconductors, e.g., HgCr$_2$Se$_4$,\cite{OsViSa98} CdCr$_2$Se$_4$,
\cite{PaHaMa02} CdMnGeP$_2$,\cite{MeIsNi00} and most promising ZnO
and GaMnN.\cite{LiYuMo05, ThOvGi02,PeAbNo03,PeAbTh03} Many
experimental results suggest room temperature (RT) ferromagnetism
in transition metal doped GaN and ZnO. However, it is still
controversial if the observed RT-ferromagnetism is intrinsic or due
to some non-resolved precipitates. Several different mechanisms \cite{LiYuMo05, JuSiMa06}
have been proposed theoretically to be
responsible for the observed magnetic hysteresis. 
A thorough discussion of
these issues is given in the review papers \cite{LiYuMo05, PeAbNo03, JuSiMa06}
and references herein.

For the functioning of our proposed ferromagnetic MOBILE we need a
conduction band spin splitting of the order of tens of meV, regardless by
which mechanism this exchange splitting is induced. In GaMnN it is
generally believed that the exchange splitting of the conduction
band is about 30-50 meV.\cite{Pearton_privcomm, LiKiLe06} In experiments the
ferromagnetic order sustains for dopings up to a few $10^{18}$ cm$^{-3}$,
in thin layers of a few nm
width.\cite{PeAbTh03} It has also been shown theoretically that a small amount of
anisotropic coupling in the 2D-Heisenberg model is sufficient to
stabilize long range order at finite temperatures.\cite{PrHwDa05}
Two-barrier RTDs based on n-GaMnN have already been investigated theoretically.\cite{LiKiLe06}

To be specific, we perform all numerical simulations for a GaMnN quantum well. The
circuit diagram of the ferromagnetic MOBILE, the material
composition, and the conduction band of the driver RTD is
schematically shown in Fig.~\ref{fig:scheme}. We consider a
two-barrier semiconductor heterostructure,
GaMnN/Al$_{1-x}$Ga$_x$N/GaMnN/Al$_{1-x}$Ga$_x$N/GaN, where a Ga
concentration of about $x=17\%$ is assumed in the barriers yielding a
barrier height of about 200 meV.\cite{AmMaMi02} The QW is undoped, whereas 
the leads consist of 15 nm long $n$-doped layers, with $n =2.78\times 10^{18}$ cm$^{-3}$ in the magnetic emitter and
$n =2.26\times10^{18}$ cm$^{-3}$ in the collector lead corresponding to a Fermi energy of
$E_f = 25$ meV at the lattice temperature of $T = 100$ K.
The magnetization $\mathbf{M}$ of the lead is considered to be
fixed, whereas the ferromagnetic QW is ``soft'', which means that its
magnetization direction can be altered by an external magnetic
field.

In order to investigate the IV-characteristics of such a structure
we follow the classic treatments of transport in nonmagnetic RTD,
\cite{VaLeLo83, CaMcDa87, Po89, OhInMu86, LaKlBo97} where coherent transport in the
whole active device region (here the barriers and the QW) is
assumed. Exploiting the symmetry of the Hamiltonian due to
translations in the plane perpendicular to the growth direction $z$
of the heterostructure, the Schr\"odinger equation can be reduced to
a one-dimensional problem. By assigning the spin quantization axis
to the fixed magnetization axis of the emitter lead the spinor
scattering states $\psi_{\sigma,E}^{i\sigma'}(z)$ with the spin
quantum number $\sigma = \pm 1/2, (\uparrow, \downarrow)$ regarding
to the boundary condition of an incident plane wave form electron
from lead $i$ [= left (L), right (R)] with spin $\sigma'$ are then
determined in the effective mass envelope function approach by
\begin{equation}\label{eq:spinor}
\left[-\frac{\hbar^2}{2}\frac{\mathrm{d}}{\mathrm{d}
z}\frac{1}{m(z)}\frac{\mathrm{d}}{\mathrm{d}
z}+U_\sigma(z)\right]\psi_\sigma^{i\sigma'}(z) = E \psi_\sigma^{i\sigma'}(z),
\end{equation}
with
\begin{equation}
U_\sigma(z) =
E_c(z)-e\phi(z)+\frac{\Delta_\mathrm{ex}(z)}{2}\boldsymbol{\sigma}\cdot\mathbf{e}_M(z).\nonumber
\end{equation}
Here, $m$ denotes the
effective electron mass, and $E$ is the total longitudinal energy of
the electron (the sum of the potential and longitudinal kinetic
energy), $E_c(z)$ denotes the intrinsic conduction band profile of
the heterostructure, $e$ is the elementary charge, $\phi$ the
electrostatic potential, $\Delta_\mathrm{ex}$ denotes the exchange
splitting of the conduction band, $\boldsymbol{\sigma}$ is the Pauli
matrices vector, and $\mathbf{e}_M$ is the unit vector of the
magnetization. For a realistic simulations space charge effects have
to be taken into account. The electrostatic potential $\phi$ is
obtained from the Poisson equation,
\begin{equation}\label{eq:poi}
\frac{\mathrm{d}}{\mathrm{d}
z}\epsilon(z)\frac{\mathrm{d}}{\mathrm{d}z} \phi(z) =
\frac{e}{\epsilon_0}\left[n(z)-N_d(z)\right],
\end{equation}
where $\epsilon$ denotes the static dielectric constant, $\epsilon_0$ is
the permeability of the vacuum, $N_d(z)$ is
the fixed donor density profile of the device, and $n(z)$ is the
electron density. The Poisson equation (\ref{eq:poi}) has to be
solved together with the Schr\"odinger equation (\ref{eq:spinor}) in
a selfconsistent way, since the quantum electron density is given
by
\begin{equation}\label{eq:quann}
n(z)
=\frac{1}{4\pi}\sum_{i,\sigma,\sigma'}\int_{U_{i,\sigma'}}^\infty
\mathrm{d}\varepsilon\:f_i(\varepsilon)
\left|\psi_{\sigma,\varepsilon}^{i,\sigma'}(z)\right|^2\frac{1}{\hbar
v_{i,\sigma'}}, \end{equation} where
\begin{equation}
 f_i(\varepsilon) = \frac{m}{\pi\hbar^2}{k_B T}
\ln\left(1+\exp(\mu_i-\varepsilon)\right).
\end{equation}
Here, $U_{i,\sigma'}$ and $ v_{i,\sigma'}$ denote the spin-dependent
potential energy and longitudinal group velocity of the electron in
the left and right lead, respectively, $\mu_i$ with $\mu_R = \mu_L
-e V_a$ is the chemical potential, where we assume that a voltage
$V_a$ is applied to the right lead, and finally $k_B$ labels the
Boltzmann constant and $T$ the lattice temperature of the leads.
Since we are interested in the generic properties of the proposed structure
we neglect the effects of polarization charges at
the interfaces, which appear in GaN due to spontaneous and
piezoelectric polarization.\cite{AmMaMi02}

After obtaining the selfconsistent potential profile the current density
is calculated in the framework of the Landauer-B\"uttiker formalism.
By assuming parabolic bands and using the same effective mass for all
layers of the heterostructure, the transmission matrix $T_{\sigma'\sigma}$
does not dependent on the transversal kinetic energy of the electrons.
Hence, the current density $j_{\sigma'\sigma}^{i\to j}$ regarding to electrons
which are incident with spin $\sigma$ from lead $i$ and end up in
lead $j$ with spin $\sigma'$ can be obtained by generalizing the Tsu-Esaki formula
\cite{TsEs73}
\begin{equation}\label{eq:j}
j_{\sigma'\sigma}^{i\to j} = \frac{e m k_B
T}{(2\pi)^2\hbar^3}
\int\limits_{\max(U_{i,\sigma},U_{j,\sigma'})}^\infty\mathrm{d}\varepsilon\:
f_i(\varepsilon)T_{\sigma'\sigma}^{i\to j}.
\end{equation}
According to time-reversal symmetry $T_{\sigma'\sigma}^{i\to j} =
T_{\sigma\sigma'}^{j\to i} $and the total current density is given by
\begin{equation}\label{eq:jtot}
j = \sum_{\sigma\sigma'} (j_{\sigma'\sigma}^{L\to R}-
j_{\sigma\sigma'}^{R\to L}),
\end{equation}
the difference of left and right flowing currents.

\section{\label{simres} Numerics and simulation results}

Following Ref. \onlinecite{VaLeLo83}, we numerically calculate the spinor scattering states in Eq.~(\ref{eq:spinor})
by applying the 4th-order Runge-Kutta (RK) scheme.
Due to the forming of quasibound states in the QW
the local density of states (LDOS) is strongly energy dependent.
Therefore, we use an adaptive energy mesh for the numerical calculation of the electron
density by rewriting the quadrature, Eq.~(\ref{eq:quann}), into an initial value problem of an
ordinary differential equation and solve it again by applying a 4th-order RK scheme.
In this way the computational costs of the numerical integration
are strongly reduced (usually 400 energy points are necessary for
a relative accuracy of $10^{-4}$, compared to 1200 grid points for an uniform energy mesh).
To achieve fast convergence (usually 5-10 iterations steps) between the Schr\"odinger equations
(\ref{eq:spinor}) and the Poisson equation (\ref{eq:poi}), we apply a predictor-corrector method.\cite{TrGaPa97}
The guess for the uniformly discretized ($\Delta z = 0.1$ nm)  electrostatic potential is
obtained by a Newton-Raphson method, where the necessary Jacobian is
estimated by using the semiclassical Thomas-Fermi approximation for the electron density.\cite{LaKlBo97}
To save computational costs the particle density in the leads $n_i(z)$ is calculated semiclassically\cite{OhInMu86}
by
\begin{equation}
n_i(z) = \frac{N_c}{2}\sum_\sigma {\mathcal F}_{1/2}
\left[\frac{\mu_i-U_\sigma(z))}{k_B T}\right],
\end{equation}
where $N_c$ is the effective conduction band density of states and $\mathcal{F}_{1/2}$ is the Fermi-Dirac
integral of order $1/2$. The transmission matrix $T_{\sigma,\sigma'}$ is obtained by the transfer-matrix technique.\cite{VaLeLo83}
Since the
transmission functions are usually ``spiky'' we use an adaptive Gauss-Kronrod scheme for an efficient
numerical calculation of the current density in Eq.~(\ref{eq:j}).

For all simulations we set the Fermi energy to $E_f = 25$ meV (all energies are measured from
the unsplit emitter conduction band edge), the lattice temperature to $T = 100$ K, 
and we use the same exchange splitting
in the QW and the emitter lead, 
$\Delta_\mathrm{ex} = 40$ meV. We assume the spin up conduction band edge to lie energetically 
higher than the spin down one,
which leads to a particle spin polarization of about -77.6 \% in the emitter lead.
An effective electron mass of $m/m_0= 0.228$ \cite{AmMaMi02} ($m_0$ denotes the free electron mass)
is used and the static dielectric
constant is set to $\epsilon = 9.5$.\cite{Brennan:book02}

\begin{figure}
\centerline{\psfig{file=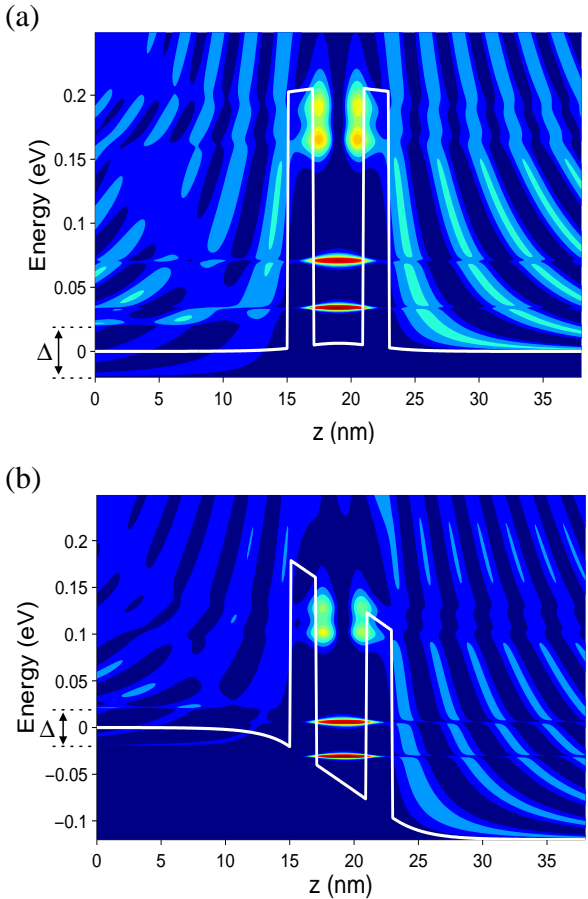,width=1\linewidth}}
\caption{(Color online) Contour plots of the local density of states versus energy and growth direction $z$
for parallel magnetization alignment in the case of (a) equilibrium $V_a = 0$ and (b) at the peak voltage
of $V_a = 0.12$ V. The exchange splitting of spin up and down level is clearly visible for the quasibound 
ground state in the well.
The solid lines indicate the self-consistent conduction band profile.  }\label{fig:LDOS}
\end{figure}

\begin{figure}
\centerline{\psfig{file=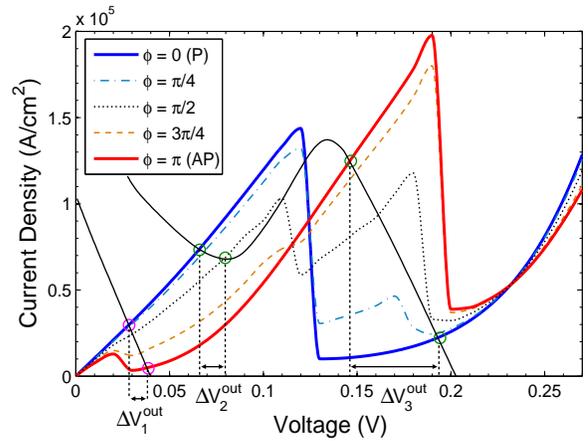,width=1\linewidth}}
\caption{(Color online) Selfconsistent current-voltage characteristics of the magnetic driver-RTD for several relative
orientations of the quantum well magnetization (indicated by the angle $\varphi$) at the temperature of $T = 100$ K.
The solid black lines show the mirrored IV curve of the load-RTD for low and high input voltages, respectively;
working points are indicated by circles.
For these fixed low and high input voltages the output voltages are restricted to the intervals $\Delta V_i^{\mathrm{out}}, i = 1,2,3$.
}\label{fig:IVsc}
\end{figure}

\begin{figure}
\centerline{\psfig{file=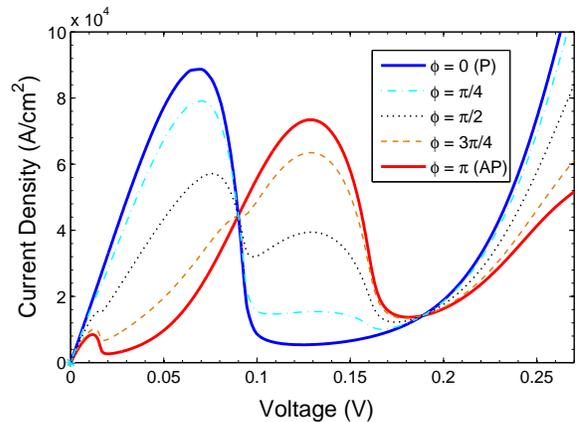,width=1\linewidth,height=5.6cm}}
\caption{(Color online) Current-voltage characteristics of the magnetic driver-RTD in the non-selfconsistent
case for different relative orientations $\varphi$ of the quantum well magnetization at the
lattice temperature $T = 100$ K. The peak voltage and currents are smaller when compared to the selfconsistent case of
Fig.~\ref{fig:IVsc} and
there exists a common crossing points for all IV-curves.     }\label{fig:IVnonsc}
\end{figure}

The contour plots of the local density of states (LDOS) of the conduction electrons  for P
magnetizations alignment in the considered
heterostructure (see Fig.~\ref{fig:scheme}) are shown in
Fig.~\ref{fig:LDOS}(a) in the equilibrium case ($V_a = 0$) and (b) at the peak voltage of $V_a = 0.12$ V.
The forming of the exchange split spin up and down quasibound states is apparent.
In the shown case of P alignment the spin up level lies energetically higher than the spin down one.
The broadening of the first spin resolved quasibound states [at about 34 meV and 71 meV in 
Fig.~\ref{fig:LDOS}(a)] in energy is
much smaller than the exchange splitting, whereas for the 
next higher quasibound states (at about 164 meV and 190 meV), which have
a node in the middle,  both spin levels already overlap. The solid lines
indicate the selfconsistent conduction band profile. In the selfconsistent case parts of
the applied voltage already drop before the first barrier but also beyond the second one as can be seen in
Fig.~\ref{fig:LDOS}(b). This is in contrast to the non-selfconsistent case, where the applied voltage is assumed to
drop linearly only in the active device region, i.e., in the QW and two barriers.

The selfconsistent IV-characteristics for the structure with the lateral dimensions indicated in Fig.~\ref{fig:scheme}
in the case of different relative orientations
of the QW-magnetization (characterized by the angle $\varphi$) is shown in Fig.~\ref{fig:IVsc}.
For comparison, Fig.~\ref{fig:IVnonsc} displays the IV-curves for the non-selfconsistent case. 
Interestingly, in the latter case there exists
a single crossing point for all IV's with the same current $I_*$ at some voltage $V_*$.
This fact suggests that the IV-characteristic
for a specific angle $\varphi$ might be written as a simple linear combination of the P and AP IV's: $I_\varphi(V) =
a(\varphi)I_0(V)+b(\varphi)I_\pi(V)$ with $a+b=1$.

The crossing in the non-selfconsistent case follows from the linearity of
the Schr\"odinger equation. For the purpose of the proof let us introduce a coordinate system, which we call the 
``0''-system, where the $z$-axis coincides with
the magnetization direction of the QW-magnetization and the $y$-axis is given by the growth direction of the heterostructure.
It is evident from the symmetry of the structure that instead of twisting the QW-magnetization by some angle $-\varphi$
it is equivalent to assume a fixed magnetization direction in the QW and to rotate the
emitter's magnetization by the angle $\varphi$. In the latter case only the boundary conditions for the
Schr\"odinger equation (\ref{eq:spinor}) are changed when compared to the P alignment.
The spinor wave functions in the emitter lead are of the
plane-wave form, when they are represented in a coordinate system (the ``$\varphi$''-system) where
the $z$-axis is given by the emitter's magnetization direction. The ``$\varphi$''-system results from
the ``$0$''-system by rotating the latter around the $y$-axis by the angle $\varphi$. Hence, the spinor
representations in the ``0''- and
``$\varphi$''-system are connected by $\{\psi\}_0 = D(-\varphi) \{\psi\}_\varphi$, where
\begin{equation}
D(\varphi)= \left(
\begin{array}{cc}
\cos(\varphi/2)&\sin(\varphi/2)\\
-\sin(\varphi/2)& \cos(\varphi/2)
\end{array}\right).
\end{equation}
Due to the linearity of the Schr\"odinger equation (\ref{eq:spinor})
and since $D(-\varphi) = \cos(\varphi/2)D(0)-\sin(\varphi/2)D(\pi)$  we can ``divide'' the problem
into finding the solution for the two boundary conditions, $\{\psi\}_0 = \cos(\varphi/2)\{\psi\}_\varphi$ and
$\{\psi\}_0 = -\sin(\varphi/2)D(\pi)\{\psi\}_\varphi$, respectively. Representing the latter in the ``$\pi$''-
coordinate system with its $+z$-axis along the $-z$-axis of the ``0''-system results in
$\{\psi\}_\pi = \sin(\varphi/2)\{\psi\}_\varphi$. Thus, we can express the amplitude for the transmission of
an incident electron with its spin aligned along the $z$-axis of the ``$\varphi$''-system
denoted by $|z_\varphi,\sigma\rangle$
to a right
moving plane wave state on the collector side with a spin eigenstate of the
$z$-axis of the ``0''-system $|z_0,\sigma'\rangle$  as
$\langle z_0, \sigma'|z_\varphi,\sigma\rangle = \cos (\varphi/2)\langle z_0, \sigma'|z_0,\sigma\rangle+
\sin(\varphi/2)\langle z_0, \sigma'|z_\pi,\sigma\rangle$. Since there is no spin precession for the
P and AP case when the electron transmits the device region and since we assume that spin flipping
scattering processes do not occur, the off-diagonal matrix elements of the
amplitude matrix $\langle z_0, \sigma'|z_0,\sigma\rangle$ vanish, whereas
$\langle z_0, \sigma'|z_\pi,\sigma\rangle = 0$ for the diagonal elements.
With this the spin-dependent transmission function, which is proportional to the squared
transmission amplitude, can be written as
\begin{eqnarray}
\lefteqn{T_{\sigma',\sigma}^\varphi\propto\left|\langle z_0, \sigma'|z_\varphi,\sigma\rangle\right|^2 =} \nonumber\\
&&\left(
\begin{array}{cc}
\cos^2(\frac{\varphi}{2})|\langle z_0, \uparrow|z_0,\uparrow\rangle|^2&\sin^2(\frac{\varphi}{2})
|\langle z_0, \uparrow|z_\pi,\downarrow\rangle|^2\\
\sin^2(\frac{\varphi}{2})|\langle z_0, \downarrow|z_\pi,\uparrow\rangle|^2&
\cos^2(\frac{\varphi}{2})|\langle z_0, \downarrow|z_0,\downarrow\rangle|^2
\end{array}\right).\nonumber
\end{eqnarray}
Using Eqs.~(\ref{eq:j}) and (\ref{eq:jtot}) finally yields the desired result,
\begin{equation}
j_\varphi(V) = \cos^2(\varphi/2)j_0(V)+\sin^2(\varphi/2)j_\pi(V).
\end{equation}
This relation is only valid in the non-selfconsistent case, since by including
space charge effects, the selfconsistent electrostatic potential will become in general
$\varphi$-dependent and the above given considerations break down. We propose that the deviation from
a common crossing point $(I_*,V_*)$ for all IV's might be used as a
criterion of the relative importance of space charge effects, modulated by $\varphi$, in the device.
As can be seen in Fig.~\ref{fig:IVsc} no common crossing
point appears in the selfconsistent case. Since the applied voltage drops over a longer spatial
region than in the non-selfconsistent case, as can be seen in Fig.~\ref{fig:LDOS}(b),
the selfconsistent peak voltages and currents are considerably
higher. Hence, for our specific device setup
space charge effects strongly influence the obtained IV-characteristics.

\begin{figure}
\centerline{\psfig{file=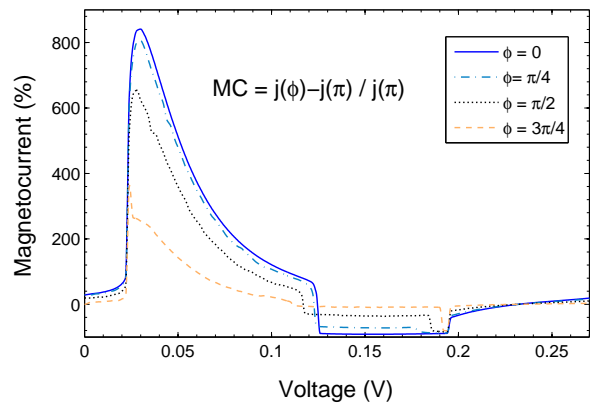,width=1\linewidth}}
\caption{(Color online) Voltage-dependent magnetocurrent (MC) in the case of different relative orientations $\varphi$
of the magnetization in the quantum well at the temperature of $T = 100$ K.}\label{fig:MC}
\end{figure}

\begin{figure}
\centerline{\psfig{file=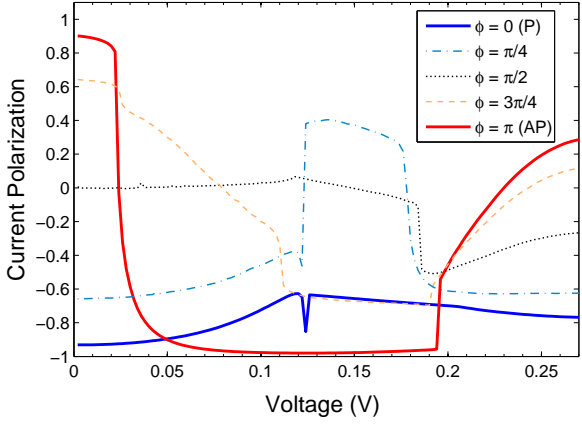,width=1\linewidth}}
\caption{(Color online) Current spin polarization as a function of the applied voltage for different orientations
of the quantum well magnetization (indicated by angle $\varphi$) at the lattice temperature of $T = 100$ K.
 }\label{fig:P}
\end{figure}

The magnetocurrent (MC) for a particular magnetization orientation $\varphi$ can be
defined as follows: 
\begin{equation}
\mathrm{MC}(\varphi) = \frac{j(\varphi)-j(\pi)}{j(\pi)}.
\end{equation}
Our selfconsistent simulations reveal that the MC increases up to high values of about 800\% 
(in the case of $\varphi = \pi$)
for voltages right after the first NDR region of the AP IV-curve ($V > 25 $ mV, see Fig.~\ref{fig:IVsc}).
Moreover, the current spin polarization with respect to the emitter spin quantization axis is given by
\begin{equation}
P_j  = \frac{j_{\uparrow\uparrow} + j_{\uparrow\downarrow}-j_{\downarrow\uparrow}-j_{\downarrow\downarrow}}
{j_{\uparrow\uparrow} + j_{\uparrow\downarrow}+j_{\downarrow\uparrow}+j_{\downarrow\downarrow}}. 
\end{equation}
As shown in
Fig.~\ref{fig:P} $P_j$
is strongly modulated when the QW magnetization is flipped from P to AP. At low voltages the
polarization can be continuously changed from -93\% for the P alignment up to +90\% for the AP orientation.
Hence, in the low voltage regime the ferromagnetic QW acts as a spin aligner.
Interestingly, in the AP case the current polarization sharply slopes from +90\% to -98\% over the
small first NDR voltage interval, which allows to use the device also as a voltage controlled spin switcher.

\begin{figure}
\centerline{\psfig{file=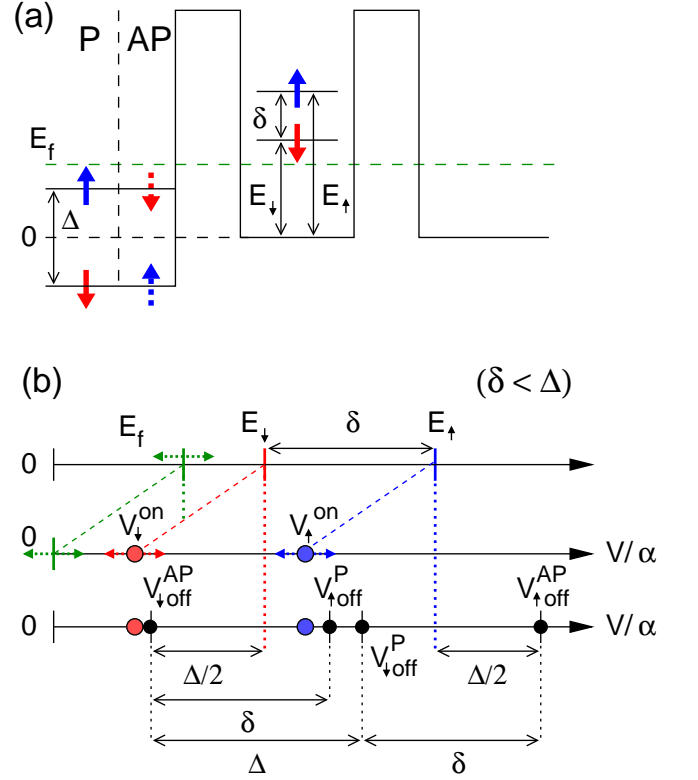,width=1\linewidth}}
\caption{(Color online) Illustration of the analytic model for the case of parallel (P) and
antiparallel (AP) magnetization alignment.
In the model the magnetization of the emitter is ``soft'', whereas the QW-magnetization is assumed to be fixed.
Part (a) shows the spin resolved conduction band profile in the emitter, 
the first quasibound states $E_\uparrow$ and $E_\downarrow$ in the quantum well, 
and the relative position of the Fermi energy level $E_f$. Part (b) displays the relative
distances of the ``on'' and ``off''-switching voltages of the spin-resolved components of the current
on a $1/\alpha$ voltage scale. The relative
position between the ``on'' and ``off'' voltages (black filled circles) can 
be changed by moving the Fermi energy level.}\label{fig:simpmod}
\end{figure}

To illustrate our results we introduce a model,
which allows to give a qualitative estimate
of the IV-curves for the extreme cases of parallel (P) $\varphi = 0$ and antiparallel (AP) $\varphi = \pi$
magnetization orientation. In particular, the model allows for a
better understanding for the influence of the following three simulation parameters:
(i) the energy difference
 for the lowest spin up and down quasi bound states $\delta= E_\uparrow-E_\downarrow$
 (it should be noted that $\delta = 37$ meV is a bit smaller than the
exchange splitting of the conduction band in the QW ($\Delta_\mathrm{ex}= 40$ meV) due to the finite barrier height),
(ii) the exchange splitting $\Delta$ in the emitter lead,
and (iii) the relative position of
the Fermi energy level $E_f$ in the emitter conduction band.
By applying a voltage to the collector lead the quasibound states $E_\sigma$ are shifted to lower energies.
At first glance the voltage dependence of quasibound energy levels can be described by
$E_\sigma(V) = E_\sigma(0)-\alpha V$, where the parameter $\alpha$ is assumed to be voltage-independent ($\alpha = 1/2$ for the
linear voltage drop and usually $\alpha < 1/2$ for the selfconsistent case).
For the following discussion it is more convenient to consider that the magnetization of the QW is fixed and, hence,
in the AP case the magnetization of the emitter is flipped. This leads to completely identical IV-characteristics since
in the AP case only the sign of the
spin polarization of the
current is changed but not its magnitude.
The model is schematically illustrated in Fig.~\ref{fig:simpmod}.

At low temperatures current can flow only if the quasibound states are dropped below the emitter Fermi energy level,
hence, the spin up and down currents are switched on at the voltages $V^{\mathrm{on}}_\sigma = \frac{1}{\alpha}(E_\sigma - E_f)$.
These spin-polarized currents are switched off, if the corresponding quasibound states are shifted
below the emitter spin up and down conduction band edges, which leads to
$V^{\mathrm{P}}_{\sigma,\mathrm{off}} = \frac{1}{\alpha}(E_\sigma - \sigma \Delta/2)$ for the P alignment and
$V^{\mathrm{AP}}_{\sigma,\mathrm{off}} = \frac{1}{\alpha}(E_\sigma + \sigma \Delta/2)$ for the AP case, respectively.
For higher temperatures the ``switching on'' relations will be thermally smeared on the order of a few $k_B T$ around the Fermi energy
but the ``switching off'' relations are still valid, as long as inelastic scattering can be neglected
as assumed in our coherent transport model. From these relations it immediately follows
that $V_{\downarrow,\mathrm{off}}^{\mathrm{AP}} < V_{\uparrow,\mathrm{off}}^{\mathrm{P}}
\leq V_{\downarrow,\mathrm{off}}^{\mathrm{P}} < V_{\uparrow,\mathrm{off}}^{\mathrm{AP}}$ for $\delta \leq \Delta$
and $V_{\downarrow,\mathrm{off}}^{\mathrm{AP}} < V_{\downarrow,\mathrm{off}}^{\mathrm{P}}
<V_{\uparrow,\mathrm{off}}^{\mathrm{P}} < V_{\uparrow,\mathrm{off}}^{\mathrm{AP}}$ for $\delta > \Delta$,
respectively. In the case of $\delta \leq \Delta$ the distances between these ``off-switching'' voltages are illustrated
in Fig.~\ref{fig:simpmod}(b); for $\delta > \Delta$ they are just given by interchanging ($\delta \leftrightarrow \Delta$).

First of all this model reveals that the peak voltage for the AP alignment is always
higher than in the P case, since $V_{\uparrow,\mathrm{off}}^{\mathrm{AP}}> V_{\sigma,\mathrm{off}}^{\mathrm{P}}$,
as obtained in our simulated IV-curves Fig.~\ref{fig:IVsc}.
For the special case of $\delta = \Delta$, which is approximately fulfilled in our simulations,
the spin up and down current in the P case can be switched off at the same voltage.
The relative position of the ``on-switching'' voltages on the voltage
scale [indicated by the colored circles in Fig.~\ref{fig:simpmod}(b)]
can be changed by the position of the Fermi energy level. By an appropriate choice of $E_f$, as
indicated in Fig.~\ref{fig:simpmod}(b), in the AP case the
spin down current can be switched off {\em before} the spin up current is switched on. This leads
to the NDR behavior in the low voltage regime, and the sharp current polarization drop.
For the P alignment the total current is in the whole voltage range dominated by the spin down
component. At low voltages up to $40$ mV almost all current ``flows'' through the spin down channel
but then also the spin up components starts to contribute to the total current, diminishing the
spin current polarization. Since both spin currents are ``switched off'' at almost the same voltage
in the P case the IV-curve is nearly ohmic in the low voltage regime.
This simple discussion shows that rotating the QW magnetization can drastically change the IV-curve.


For the ``intermediate'' case of perpendicular orientation $(\varphi = \pi/2)$ (see Fig.~\ref{fig:IVsc}) 
two peak voltages appear
with nearly the same values as those obtained for the P and AP alignment, respectively.
However, most important for observing DMR is that the peak current is already remarkably reduced when the
relative magnetization orientation is tilted, say by an angle of a few tens of degrees, out of the P alignment.
If we assume the load peak current is smaller than the driver peak current in the P case,
the mono-to-bistable transition results in a low output voltage. However, if we tilt the magnetization
orientation, the load and driver peak current become equal at some angle $\varphi_\mathrm{th}$, which
can be called the ``threshold angle''. For $\varphi > \varphi_{\mathrm{th}}$ a high output voltage is obtained in the
bistable regime. Hence, the output voltage suddenly jumps from low to high after performing the mono-to-bistable
transition, effectively realizing DMR. The threshold angle can be controlled indirectly by
properly tuning the load peak current, which is altered by an external gate voltage applied to the load device.
Assuming a fixed low and high input voltage the output voltage is restricted to three different
voltage intervals $\Delta V_i^{\mathrm{out}}, i=1,2,3$ as illustrated in Fig.~\ref{fig:IVsc}.
The high voltage
interval $\Delta V_3^{\mathrm{out}}$ is considerably separated from
the low voltages intervals $\Delta V_1^{\mathrm{out}}$ and $\Delta
V_2^{\mathrm{out}}$. This allows for a direct digital detection of
the tilted QW magnetization.

The question of how fast the mono-to-bistable transition can be performed is closely connected to the
subtle and still not clarified
problem of determining the error rates in MOBILEs. The tilting of the QW magnetization leads
to a redistribution of the quasibound states, which usually takes place on the time scale of the order of a 
hundred of femtoseconds. The switching time of RTDs, however, is limited by the ``classical'' RC time constant,
which is typically of the order of a few picoseconds.\cite{DiOeRo89}
In experiments the conventional MOBILEs randomly jump between high and low output voltage
in the transition region
and an erroneous transition can occur due to parasitic 
capacitances or external electrical noise.\cite{Maezawa_privcomm}
Transient studies of conventional
MOBILEs based on an equivalent circuit model have shown that an error-free transitions with clock rise
times on the order of the RC time of the RTD are possible if the output capacitance
$C_\mathrm{out} < (k-1) C_\mathrm{RTD}$,\cite{Ma95} where $C_\mathrm{RTD}$ is an average capacitance of the RTD and $k$
is the ratio of load to driver peak current.  Recently, conventional MOBILEs have been
demonstrated to work up to frequencies as high as
100 GHz by employing a symmetric clock configuration for the input voltage.\cite{MaSuKi06}
This gives reason for a possible application of the proposed ferromagnetic MOBILE as a very fast ``readout''
of magnetically stored information.

\begin{figure}
\centerline{\psfig{file=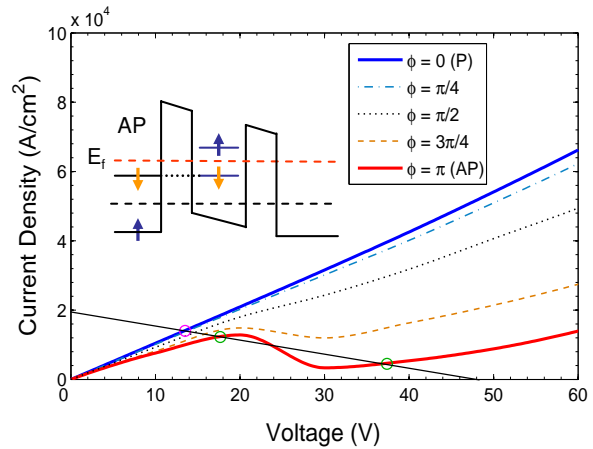,width=1\linewidth}}
\caption{(Color online) Blow up of the selfconsistent current-voltage characteristics
displayed in Fig.~\ref{fig:IVsc} for low voltages. By replacing the load-RTD by a linear load
resistance (indicated
by the black solid line) allows to perform the mono-to-bistable transition by flipping the QW magnetization from
P to AP. The working points are indicated by circles. The inset schematically shows the arrangement
of the quasibound levels at the peak voltage in the AP case.}\label{fig:IVblow}
\end{figure}

Fig.~\ref{fig:IVblow} shows a blow up of the selfconsistent IV-characteristics of Fig.~\ref{fig:IVsc} for low voltages.
By rotating the QW
magnetization the IV-curves change from ohmic to NDR-behavior.
As illustrated in the inset of Fig.~\ref{fig:IVblow} the spin down current becomes
off resonance already before the current can flow ``through'' the spin up quasibound state
leading effectively to NDR. This interesting behavior can be used to perform
the mono-to-bistable transition by twisting the QW magnetization
instead of changing the input voltage from low to high as discussed above. For this,
we assume to use a linear load resistance instead of the load-RTD in
the circuit setup of Fig.~\ref{fig:scheme}. For an appropriately high input voltage
two stable working points are obtained in the AP case, whereas only one crossing point appear in the load line
diagram for the P
alignment (see Fig.~\ref{fig:IVblow}). With this again DMR can be realized as follows. Let us assume that at the
beginning the circuit operates
at the high voltage working point of the AP orientation. If we tilt the QW magnetization, suddenly at some
threshold angle, the circuit is switched from the bistable to the monostable regime, leading to a
discrete jump from high to low output voltage. This allows to detect ``digitally'' a disturbance of the AP
magnetization alignment. After this detection and after the 
recovering of the AP orientation the circuit will end up in
the low voltage state of the bistable regime. By applying a small current pulse to the circuit it
can again be reset to
the initial high voltage state. Another application could be 
a memory cell, in which the binary information is stored in the P and AP configuration and
is read out by a small current pulse, which leads in the AP case to a voltage swing to the
high voltage state.

\section{\label{conclus}Conclusions}

We have proposed a ferromagnetic MOBILE, where the driver-RTD comprises a ferromagnetic emitter
and QW. By using a selfconsistent coherent transport model we have shown that by changing
the relative orientation of the two magnetizations the
IV-characteristics are strongly modulated and that nonvolatile DMR can be realized with this circuit.
In particular, this allows for an electrical and direct digital detection
of a small distortion out of the P alignment. The comparison of the selfconsistent with the
non-selfconsistent model, where a linear voltage drop is
assumed in the device, reveals that space charge effects have to be included to get more
realistic IV-characteristics. High MCs up to $800\%$ are obtained in the AP case, at 100 K, and
the current spin polarization can be continuously changed from $+90\%$ to $-93\%$ by either
flipping the QW magnetization or by altering the applied voltage for the AP alignment.
By a proper choice of the Fermi energy level and the magnitudes of
the exchange splitting in both the emitter and QW,
respectively, the IV-curves can be changed in the low voltage regime
from ohmic to NDR behavior by twisting the QW-magnetization from P to AP alignment.
When serially connected to a load resistance this allows to accomplish
a mono-to-bistable working point transition just by flipping the
QW magnetization, in contrast to the usual way of performing the transition where the input voltage is increased
from low to high level. Since conventional MOBILEs have been demonstrated to work up to 100 GHz
the proposed device might be useful for performing very fast detections of magnetic signals and for
realizing fast magnetic random access memories.

\section*{Acknowledgment}

 This work has been supported by the Deutsche
Forschungsgemeinschaft SFB 689. The authors thank S. J. Pearton and K. Maezawa for
valuable discussions.


\end{document}